\documentclass[preprint]{ptephy_v1}

\preprintnumber{YITP-21-18, OU-HET-1089, KEK-CP-0380}

\usepackage{graphicx}
\usepackage{color}
\usepackage{bm}
\usepackage{multirow}
\usepackage{amsmath}
\usepackage{amsfonts}
\begin{document}

\newcommand{\YITP}{
  Center for Gravitational Physics, Yukawa Institute for Theoretical Physics, 
  Kyoto University, 
  Kyoto 606-8502, Japan
}
\newcommand{\Osaka}{
  Department of Physics, Osaka University,
  Toyonaka 560-0043, Japan
}
\newcommand{\Edinburgh}{
School of Physics and Astronomy, The University of Edinburgh, Edinburgh EH9 3JZ, United Kingdom 
}
\newcommand{\KEK}{
  High Energy Accelerator Research Organization (KEK), 
  Tsukuba 305-0801, Japan
}
\newcommand{\Sokendai}{
  School of High Energy Accelerator Science,
  The Graduate University for Advanced Studies (Sokendai), 
  Tsukuba 305-0801, Japan
}
\newcommand{\RIKENCCS}{
RIKEN Center for Computational Science,
7-1-26 Minatojima-minami-machi, Chuo-ku, Kobe, Hyogo 650-0047, Japan
}
\newcommand{\JAEA}{
Advanced Science Research Center, Japan Atomic Energy Agency (JAEA), Tokai 319-1195, Japan
}

\title{
  Role of axial $U(1)$ anomaly in chiral susceptibility
  of QCD at high temperature
}
\author{JLQCD Collaboration: S.~Aoki}
\affil{\YITP}
\author{Y.~Aoki}
\affil{\RIKENCCS}
\author{H.~Fukaya}
\affil{\Osaka}
\author{S.~Hashimoto}
\affil{\KEK}
\affil{\Sokendai}
\author{C.~Rohrhofer}
\affil{\Osaka}
\author{K.~Suzuki}
\affil{\JAEA}

\begin{abstract}
The chiral susceptibility, or the first derivative of
  the chiral condensate with respect to the quark mass,
  is often used as a probe for the QCD phase transition since
  the chiral condensate is an order parameter of
$SU(2)_L \times SU(2)_R$ symmetry breaking.
However, the chiral condensate also breaks the axial $U(1)$ symmetry, which is usually
not paid attention to as it is already broken by anomaly and apparently gives little impact on the transition.
We investigate the susceptibilities in the scalar 
and pseudoscalar channels
in order to quantify how much the axial $U(1)$ breaking
contributes to the chiral phase transition.
Employing a chirally symmetric lattice Dirac operator,
and its eigenmode decomposition,
we  separate the axial $U(1)$ breaking effects from others. 
Our result in two-flavor QCD indicates that
both of the connected and disconnected chiral susceptibilities are
dominated by the axial $U(1)$ breaking at temperatures $T\gtrsim 190$ MeV
after the quadratically divergent constant is subtracted.
\end{abstract}
\maketitle

\section{Introduction}
Properties of phase transition are largely governed by symmetries that
are broken/restored at the transition. In
Quantum Chromodynamics (QCD) with two degenerate
dynamical quarks (up and down), the relevant symmetry is that of
flavor rotation of left- and right-handed quark fields, {\it i.e.}
$SU(2)_L\times SU(2)_R$ chiral symmetry, which is spontaneously broken
at low temperatures but is believed to be recovered
at some high temperature our universe experienced at its early stage.
The chiral condensate $\Sigma(m)=-\sum_x\langle S^0(x)\rangle/V$, defined
with a flavor singlet scalar quark bilinear operator $S^0(x)$ and the four-volume $V$,
as well as its derivative $\chi(m)=\frac{\partial}{\partial m}\Sigma(m)$ known as
the chiral susceptibility, are often used to probe the so-called chiral phase transition
\cite{Karsch:1994hm, Aoki:2006we, Cheng:2006qk, Bazavov:2011nk, Bhattacharya:2014ara, Bonati:2015bha, Brandt:2016daq, Taniguchi:2016ofw, Ding:2019prx}.

The condensate also breaks the flavor-singlet axial symmetry $U(1)_A$ but
its relevance 
is not immediately clear, since it is broken by quantum anomaly,
which exists at any energy scale.
The $U(1)_A$ anomaly may still affect the low-energy dynamics
as it is related to the topology of the gluon field
and the zero eigenstates of the Dirac operator through the index theorem.
In fact, the founders of QCD (see \cite{Callan:1977gz} for example) strongly suggested that
the $SU(2)_L\times SU(2)_R$ breaking is triggered by the
topologically nontrivial configuration of gluons
and even quantitative estimate was made in \cite{Diakonov:1984vw}.
However, this view of the $U(1)_A$ anomaly as the origin of
the $SU(2)_L\times SU(2)_R$ breaking is not widely appreciated today
since early lattice simulations reported survival of the axial $U(1)$ anomaly
near the critical temperature, and thus, the $U(1)_A$ anomaly apparently gives little impact on the transition.

In this work, we revisit this issue, using lattice QCD with exactly
chiral and flavor symmetric quarks.
The index theorem is satisfied on the lattice to a good
precision so that the relation between topological gauge excitation and
fermion near-zero mode remains intact. By an eigenmode decomposition of
the Dirac operator and quark propagators \cite{Aoki:2012yj, Cossu:2015kfa, Tomiya:2016jwr, Aoki:2020noz}, we can
unambiguously separate
the $U(1)_A$ breaking effect from others in the
chiral susceptibility $\chi(m)=\sum_x\langle S^0(x)S^0(0)\rangle-V\langle S^0(0)\rangle^2$.
With the exact chiral symmetry, we can avoid severe lattice artifact
that can induce large overestimate of the $U(1)_A$ breaking as
demonstrated in 
\cite{Cossu:2015kfa, Tomiya:2016jwr}. 

We find that $\chi(m)$ in the high temperature phase mostly probes
the presence/absence of the $U(1)_A$ symmetry:
the connected part is dominated by the $U(1)_A$ susceptibility defined as
$\sum_x\langle P^a(x)P^a(0)-S^a(x)S^a(0)\rangle$, where $S^a(x)$ and
$P^a(x)$ are iso-triplet scalar and pseudo-scalar operators,
and the disconnected part is governed by the topological susceptibility,
which measures the instanton number variance.
Meanwhile, the $SU(2)_L\times SU(2)_R$ susceptibilities remain small 
even when the chiral condensate and 
$U(1)_A$ susceptibility become non-zero due to finite quark masses.

This result suggests a possibility that the chiral phase transition is actually driven by the
$U(1)_A$ breaking as suggested in the early stage of QCD.
In \cite{Pisarski:1983ms}, it was argued that
if the $U(1)_A$ breaking is kept large at the critical temperature
the transition is likely to be the second order.
But if the $U(1)_A$ symmetry effectively ``emerges'',
the order or the universality class of the transition differs
from the naive expectation, which would require
changes in the current understanding of the early universe.

\section{Dirac eigenmode decomposition of susceptibilities}
Let us start with the $N_f$--flavor QCD partition function
with a nonzero vacuum angle $\theta$,
\begin{eqnarray}
  Z(m,\theta) &=& \int [dA] \det(D(A)+m)^{N_f}e^{-S_G(A)+i\theta Q(A)},
\end{eqnarray}
where the path integral over the gauge field $A$
is performed with a weight given  by the gauge action $S_G(A)$,
topological charge $Q(A)$, or equivalently the index of the Dirac operator $D(A)$, 
and the fermion determinant with a degenerate quark mass $m$.
We have used a continuum notation for simplicity.
The lattice formulas in terms of
the overlap-Dirac operator \cite{Neuberger:1997fp} will be given later.

Denoting the eigenvalues of $D(A)$  by $i\lambda(A)$,
among which every nonzero mode appears in a pair with its conjugate $-i\lambda(A)$,
the chiral condensate at $\theta=0$ is decomposed as 
\begin{eqnarray}
\Sigma(m)
  = \frac{1}{N_fV}\frac{\partial}{\partial m}\ln Z(m,0)
  =\frac{1}{V}\left\langle \sum_{\lambda(A)} \frac{m}{\lambda(A)^2+m^2}\right\rangle.
\end{eqnarray}
Here and in the following, the expectation value of a quantity $X(A)$ (with $\theta=0$)
is written as $\langle X(A)\rangle$.

The chiral susceptibility, defined as a derivative of $\Sigma(m)$
with respect to $m$, may be decomposed into two parts.
The ``connected'' susceptibility $\chi^{\rm con.}(m)$ is 
a derivative of the chiral condensate with respect to the valence quark mass $m_{v}$,
while the ``disconnected'' part $\chi^{\rm dis.}(m)$ is that with respect to the sea quark mass $m_{s}$,
both with setting $m_v=m_s=m$ after all.

In the connected susceptibility, we have a term $\Sigma(m)/m$,
which is a unique source of a quadratic divergence, while other pieces
are only logarithmically divergent.
To remove this quadratic divergence, we 
introduce a subtracted condensate $\Sigma_{\rm sub.}(m)$,
\begin{eqnarray}
  \frac{\Sigma_{\rm sub.}(m)}{m} &=&
  \left[\frac{\Sigma(m)}{m}-\frac{\langle |Q(A)|\rangle}{m^2V}\right]
  \nonumber\\&&
  -
  \left[
    \frac{\Sigma(m_{\rm ref})}{m_{\rm ref}}
  -\frac{\langle |Q(A)|\rangle|_{m=m_{\rm ref}}}{m^2_{\rm ref}V}\right],
\end{eqnarray}
with a reference quark mass $m_{\rm ref}$.
The term with $|Q(A)|$  eliminates
the contribution from chiral zero modes, which
is expected to vanish in the large $V$ limit.


The connected susceptibility (with the subtraction above)
can be written as
\begin{eqnarray}
  \label{eq:con}
  \chi_{\rm sub.}^{\rm con.}(m)  &=&
  -\Delta^{\rm con.}_{U(1)}(m)+ \frac{\Sigma_{\rm sub.}(m)}{m}
  +  \frac{\langle |Q(A)|\rangle}{m^2V},
\end{eqnarray}
where 
\begin{eqnarray}
 \Delta_{U(1)}^{\rm con.}(m) = \frac{1}{V}\left\langle \sum_{\lambda(A)}\frac{2m^2}{(\lambda(A)^2+m^2)^2}\right\rangle
\end{eqnarray}
is equivalent to the axial $U(1)$ susceptibility $\sum_x \left[\langle P^a(x)P^a(0)\rangle-\langle S^a(x)S^a(0)\rangle\right]$.
(See \cite{Cossu:2015kfa, Tomiya:2016jwr} for the details).
On the other hand, the eigenvalue decomposition of the disconnected part is
\begin{eqnarray}
  \chi^{\rm dis.}(m) &=&  \frac{N_f}{V}
  \left[\left\langle \left(\sum_{\lambda(A)} \frac{m}{\lambda(A)^2+m^2}\right)^2\right\rangle
    -(\Sigma(m)V)^2
    \right].
\end{eqnarray}

From the  $\theta$ dependence of $Z(m,\theta)$, we obtain the
topological susceptibility,
\begin{eqnarray}
  \chi_t(m) = -\left.\frac{1}{V}\frac{\partial^2}{\partial \theta^2}
  \ln Z(m,\theta)\right.|_{\theta=0} =\frac{\langle Q(A)^2\rangle -\langle Q(A)\rangle^2}{V}.
\end{eqnarray}
Absorbing
the angle $\theta$ to the mass term $m\to m\exp(i\gamma_5\theta/N_f)$,
we can relate $\chi_t(m)$ to the chiral condensate and
the pseudoscalar susceptibility $\sum_x \langle P^0(x)P^0(0)\rangle$:
\begin{eqnarray}
\chi_t(m)
  = m\left.\left[\frac{\partial }{\partial \theta}\langle \bar{q}i\gamma_5 e^{i\gamma_5\theta/N_f}q\rangle_\theta\right]\right|_{\theta=0}
= - \sum_x \langle P^0(x)P^0(0)\rangle -\frac{\Sigma(m)}{m}.
\end{eqnarray}


We can now see 
that the $U(1)_A$ and  $SU(2)_L\times SU(2)_R$ symmetries are intimately related \cite{Nicola:2018vug,Nicola:2020smo}.
Two possible probes of the $SU(2)_L\times SU(2)_R$ symmetry given by
\begin{eqnarray}
  \label{eq:SU21}
  \Delta_{SU(2)}^{(1)}(m) &\equiv&  
  \sum_x \langle S^0(x)S^0(0)-P^a(x)P^a(0)\rangle-V\langle S^0(0)\rangle^2=\chi^{\rm dis.}(m)-\Delta_{U(1)}^{\rm con.}(m),\\
  \Delta_{SU(2)}^{(2)}(m) &\equiv&  
  \sum_x \langle S^a(x)S^a(0)-P^0(x)P^0(0)\rangle=\frac{N_f}{m^2}\chi_t(m)
  -\Delta_{U(1)}^{\rm con.}(m),
  \label{eq:SU22}
\end{eqnarray}
are actually written using the $U(1)_A$ related quantities $\Delta_{U(1)}^{\rm con.}(m)$, $\chi_t(m)$ and $\chi^{\rm dis.}(m)$.
When the axial $U(1)$ anomaly is active so that $\Delta_{U(1)}^{\rm con.}(m)$
\cite{Bazavov:2012qja,Cossu:2013uua,Buchoff:2013nra,Dick:2015twa,Ishikawa:2017nwl,1826587,Kaczmarek:2021ser}
is nonzero, the recovery of the $SU(2)_L\times SU(2)_R$ requires
a fine tuning 
\begin{eqnarray}
  \lim_{m\to 0} \chi^{\rm dis.}(m) = \lim_{m\to 0}\Delta_{U(1)}^{\rm con.}(m)
  = \lim_{m\to 0}\frac{N_f}{m^2}\chi_t(m),
\end{eqnarray}
which is highly nontrivial.

From Eqs.(\ref{eq:con}), (\ref{eq:SU21}) and (\ref{eq:SU22}), we can separate
the $U(1)_A$ breaking contributions $\chi^{\rm con.}_A(m)$ and $\chi^{\rm dis.}_A(m)$
from the connected and disconnected parts of
the chiral susceptibility, $\chi_{\rm sub.}^{\rm con.}(m)$ and $\chi^{\rm dis.}(m)$ respectively, as
\begin{eqnarray}
  \chi^{\rm con.}_A(m) &=& -\Delta_{U(1)}^{\rm con.}(m)+\frac{\langle |Q(A)|\rangle}{m^2V},\\
  \chi^{\rm dis.}_A(m) &=& \frac{N_f}{m^2}\chi_t(m).
\end{eqnarray}
Then the remnants are $\chi_{\rm sub.}^{\rm con.}(m)-\chi^{\rm con.}_A(m)=\Sigma_{\rm sub.}(m)/m$
and $\chi^{\rm dis.}(m)-\chi^{\rm dis.}_A(m)=\Delta_{SU(2)}^{(1)}(m)-\Delta_{SU(2)}^{(2)}(m)$, respectively.

These formulas can be promoted to those of lattice QCD with the overlap fermion \cite{Cossu:2015kfa}.
Denoting the eigenvalue of massive overlap-Dirac operator $\gamma_5((1-m)D_{\rm ov}+m)$ by $\lambda_m$,
the eigenvalue decomposition can be obtained by replacing
$\frac{1}{\lambda(A)^2+m^2}\;\;\mbox{by}\;\;\frac{(1-\lambda_m^2)}{(1-m^2)\lambda_m^2}$
(Here and in the following, we take the lattice spacing unity).
In the following we numerically study how much the $U(1)_A$-related pieces
$\chi^{\rm con./dis.}_A(m)$ dominate the signal of
the chiral susceptibility 
in $N_f=2$ lattice QCD.
We employ a lattice fermion formulation that precisely preserves chiral
symmetry, which is essential in the above formulas with spectral decomposition.

\section{Lattice simulation}
We use the gauge field ensembles generated
in \cite{Aoki:2020noz}.
We employ the tree-level improved Symanzik gauge action
and the M\"obius domain-wall fermion \cite{Brower:2005qw} action for the simulations.
We include the overlap fermion determinant utilizing
a reweighting technique in order to eliminate
systematics due to any violation of the chiral symmetry,
as well as those due to the mixed action.
The lattice spacing is fixed to $a=0.074$ fm, and
four different temperatures are chosen 
taking a set of the temporal lattice extent $L_t=8,10,12$ and 14,
which covers $190 \leq T \leq 330$ MeV.
We fix the lattice size to
$L=32$, which corresponds to 2.4 fm.
At $T=220$ MeV, three different lattice sizes
$L=24,32,40$ are taken in order to
check if the finite volume effect is under control.
The range of quark mass covers the
physical up and down quark mass, estimated to be $m$ = 0.0014(2)
from the pion mass $m_\pi=0.135(8)$ at $T=0$ and $m=0.01$.
We use the $m=0.005$, which is the highest quark mass
at $T=190$ MeV simulations, on $L=32$ lattices
as the reference point $m_{\rm ref}$ for the subtraction of
the connected chiral susceptibility.

We compute 40 lowest eigenvalues
of the massive overlap-Dirac operator,
as well as those of
four-dimensional effective operator of the
M\"obius domain-wall fermion.
For both operators, we can
identify the index $Q(A)$ as the number of isolated chiral zero modes.
At the lowest temperature, the 40th eigenvalue
is $\sim 0.08$ ($\sim 210$ MeV).

\begin{figure*}[bth]
  \centering
    \includegraphics[width=12cm]{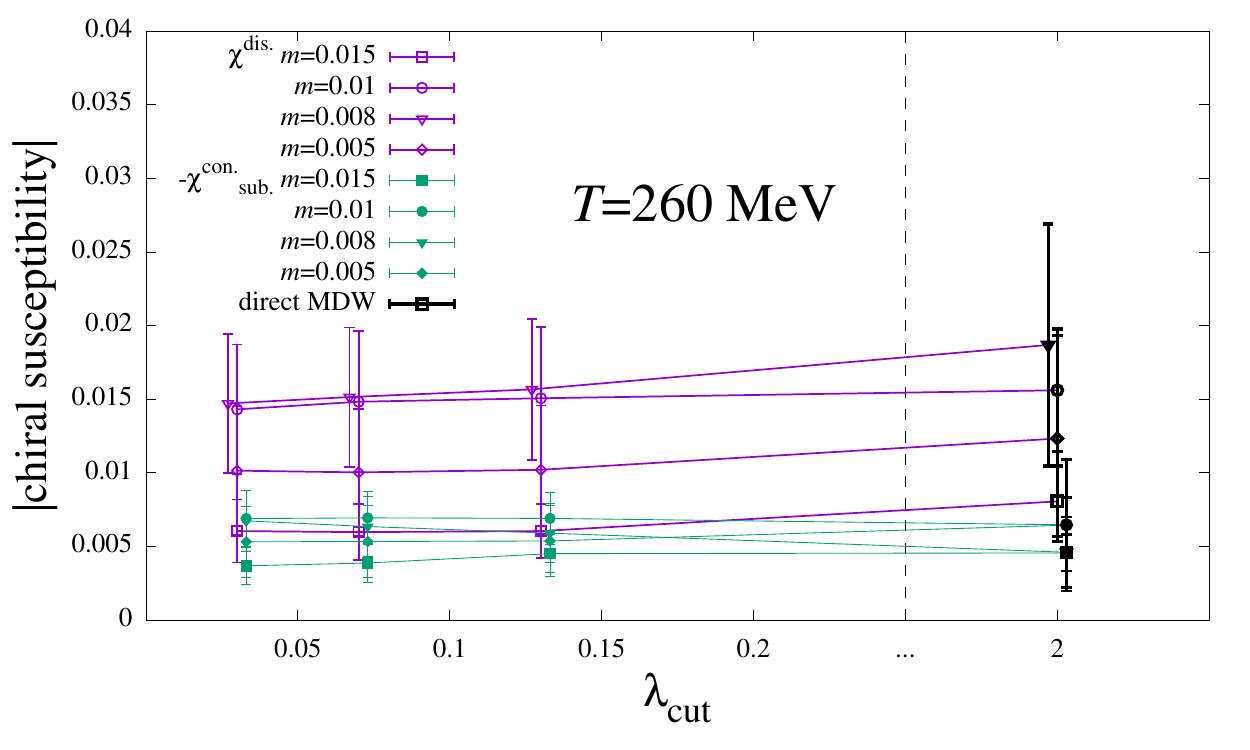}
    \includegraphics[width=12cm]{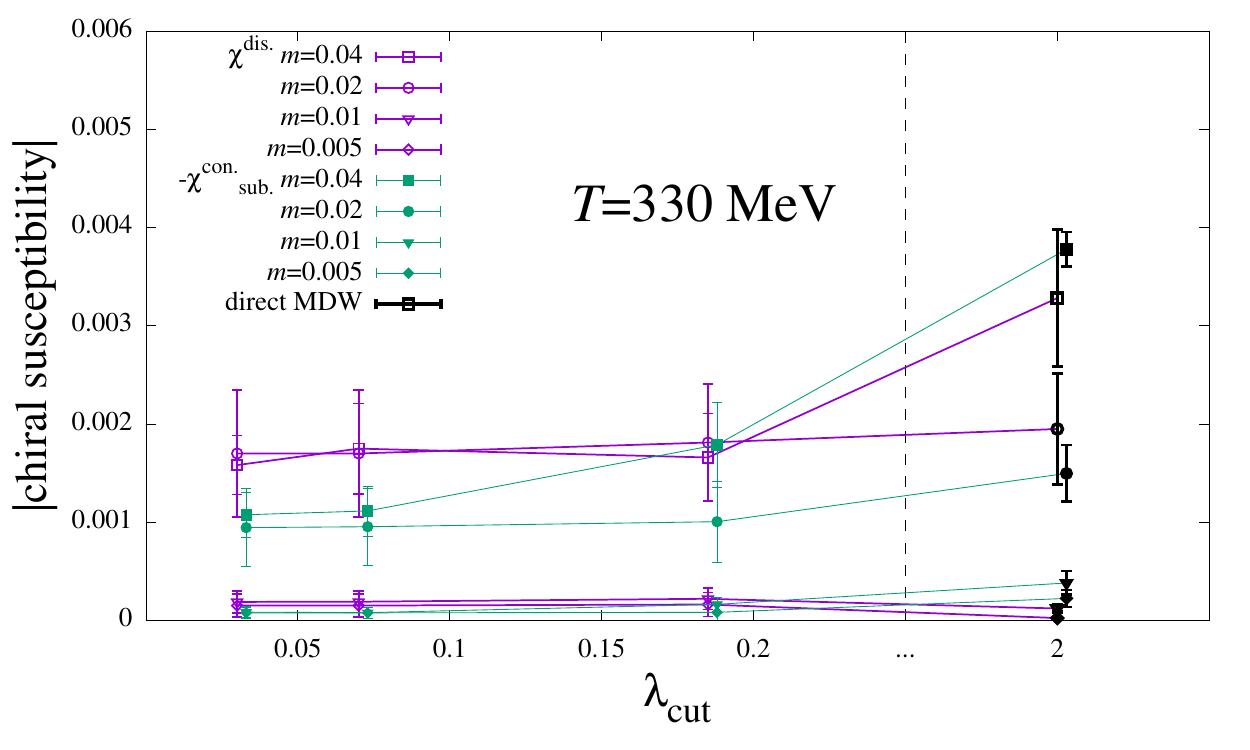}
  \caption{
    Cut-off $\lambda_{\rm cut}$ dependence of the
    chiral susceptibility at $T=260$ MeV (top panel) and $T=330$ MeV (bottom).
    The result for $\chi^{\rm dis.}$ is plotted  by open symbols,
    while that for $-\chi^{\rm con.}_{\rm sub.}$ is shown by filled symbols.
    The thick symbols plotted at the lattice cut-off $=2$
    denote those obtained from the direct inversion
    of the M\"obius domain-wall operator.
  }
  \label{fig:lowmode}
\end{figure*}

Since the number of stored eigenvalues is limited,
we truncate the summation in the spectral decomposition
of the chiral susceptibilities.
In Fig.\ref{fig:lowmode}, we plot
the cut-off dependence of the
the (subtracted) chiral susceptibilities.
We find
for $T\leq 260$ MeV,
both $\chi^{\rm con.}_{\rm sub.}$ and $\chi^{\rm dis.}$
with the reweighted overlap fermion show
a good saturation already at $\lambda=0.07$ in all the simulated ensembles.
At $T=260$ MeV, we also find a good agreement with
a full measurement without the truncation computed with the M\"obius domain-wall Dirac operator.
At $T=330$ MeV, on the other hand, the low-mode approximation does not reproduce
the full result especially at heavier quark masses.
In our previous study \cite{Aoki:2020noz} we found that
at this temperature the low-lying modes are almost absent
and the observables are insensitive to the violation of the
lattice chiral symmetry.
Therefore, in the following analysis at $T=330$ MeV,  we take the full computation with
the M\"obius domain-wall Dirac operator
and use the low-mode approximation of the overlap-Dirac fermion
at $\lambda_{\rm cut}=0.07$ ($\sim 180$ MeV) for other ensembles of $T\leq 260$ MeV.

The statistical uncertainty is estimated by the jackknife method
after binning the data in every 1000 trajectories with which
the autocorrelation is negligible.


%

\section{Numerical results}

We summarize our numerical results
in Tab.~\ref{tab:result}.

\begin{table*}[tbp]
  \centering
  \begin{tabular}{ccc|cc|cc}
    \hline\hline
    $T$(MeV) & $L^3\times L_t$ &  $m$ & $\chi_{\rm sub.}^{\rm con.}$ & $\chi_{A}^{\rm con.}$& $\chi^{\rm dis.}$& $\chi_A^{\rm dis.}$\\
    \hline
    190 & $32^3\times 14$ &  0.005 & -0.074(07) & -0.074(12)& 0.090(12)& 0.077(17)
    \\
    &  & 0.00375 & -0.101(20)& -0.106(26)& 0.144(39)& 0.189(49)
    \\
    &  & 0.0025 & -0.058(12)& -0.056(11)& 0.087(15)& 0.079(21)
    \\
    &  & 0.001 & -0.0188(64)& -0.00130(45)& 0.0020(06)& 1.6(16)e-7
    \\
    \hline
    220 & $24^3\times 12$ &  0.01 & -0.0224(23)& -0.0202(34)& 0.0399(75)& 0.0338(68)
    \\
    && 0.005 & -0.0367(87)& -0.0332(88)& 0.066(20)& 0.072(24)
    \\
    && 0.00375& -0.0081(28)& -0.0041(27)& 0.0079(52)& 0.0070(53)
    \\
    && 0.0025& -0.0125(55)& -0.0095(54)& 0.019(11)& 0.018(11)
    \\
    && 0.001& -0.0033(25)& -0.0002(01)& 0.00033(24) & 0(0)
    \\
    \cline{2-7}
    & $32^3\times 12$   & 0.01 & -0.0284(25)& -0.0325(38)& 0.044(07)& 0.049(11)
    \\
    && 0.005 & -0.0311(42)& -0.0311(48)& 0.065(11)& 0.068(14)
    \\
    && 0.00375& -0.00682(83)& -0.00270(70)& 0.0050(13)& 0.0038(13)
    \\
    && 0.0025& -0.0073(49)& -0.0062(48)& 0.0121(94)& 0.0112(95)
    \\
    && 0.001& -0.0016(12)& -0.00016(06)& 0.00030(12)& 1.8(18)e-5
    \\
    \cline{2-7}
    & $40^3\times 12$   & 0.01 & -0.0270(15)& -0.0349(28)& 0.0417(56)& 0.0397(49)
    \\
    && 0.005 & -0.0305(31)& -0.0371(56)& 0.0526(65)& 0.0433(54)
    \\
    \hline
    260 & $32^3\times 10$ &  0.015& -0.0039(13)& -0.0038(14)& 0.0060(19)& 0.0061(24)
    \\
    && 0.01 & -0.0070(18)& -0.0077(24)& 0.0148(48)& 0.0141(43)
    \\
    && 0.008& -0.0064(20)& -0.0089(32)& 0.0152(47)& 0.0117(38)
    \\
    && 0.005& -0.0054(24)& -0.0054(24)& 0.0100(43)& 0.0103(45)
    \\
     \hline
    330 & $32^3\times 8$ &  0.040& -0.00378(17)& -0.00306(21)& 0.00328(70)& 0.00219(40)
    \\
    &&  0.020&-0.00150(29)& -0.00145(30)& 0.00195(57)& 0.00148(49)
    \\
    &&  0.015&-0.00145(65)& -0.00151(82)& 0.0027(19)& 0.0017(11)
    \\
    && 0.01 &-0.000386(95)& -0.000183(62)& 0.00012(03)& 0.00044(31)
    \\
    && 0.005& -0.000222(87)& -0.000222(77)& 2.33(53)e-5& 0(0)
    \\
    && 0.001& -0.00010(10)& -2.81(58)e-5& 8.6(14)e-7&0(0)
    \\
\hline\hline
  \end{tabular}
  \caption{Summary of results}
  \label{tab:result}
\end{table*}

\begin{figure*}[bth]
  \centering
  \includegraphics[width=12cm]{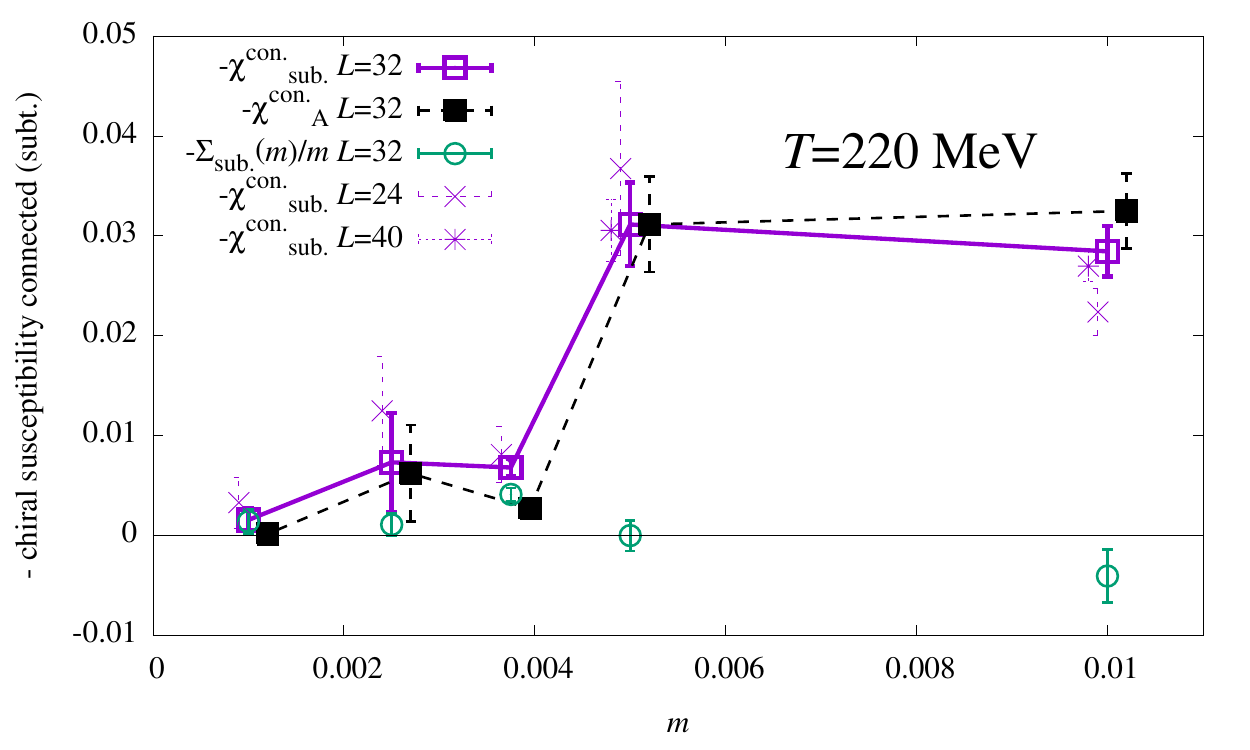}
    \includegraphics[width=12cm]{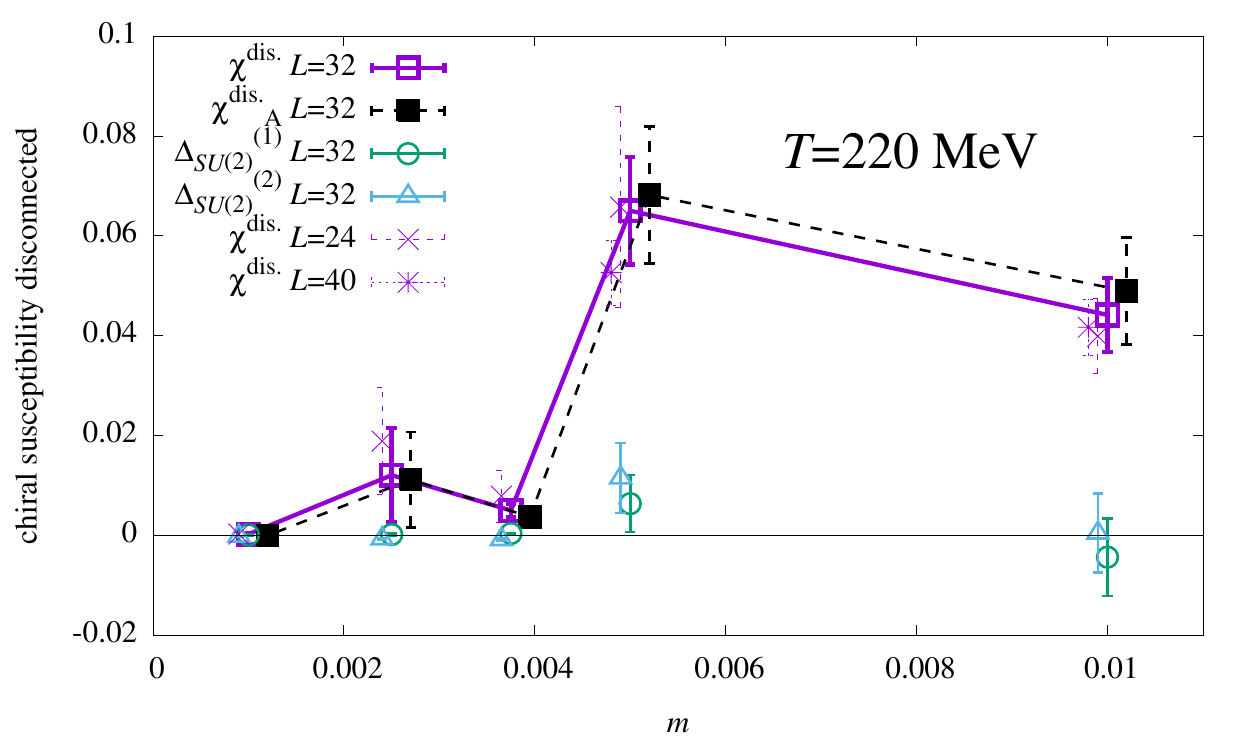}
    \caption{
      Quark mass dependence of the connected (top panel) and disconnected (bottom) 
      chiral susceptibilities on the $L=32$ lattice (open squares).
      The contribution from the axial $U(1)$ breaking (filled squares) saturates the signal,
      while the remaining $\Sigma_{\rm sub.}(m)/m$ and $\Delta_{SU(2)}^{(1,2)}(m)$ plotted by
      open circles and triangles are small.
      The $L=24$ (crosses) and $L=40$ (stars) data show no significant volume dependence.
      Note that the sign of the connected part is flipped.
  }
  \label{fig:T220}
\end{figure*}

In Fig.~\ref{fig:T220}, we present the results for
the connected part $\chi_{\rm sub.}^{\rm con.}$ (top panel)
and disconnected data $\chi^{\rm dis.}$ (bottom)
of the chiral susceptibility at $T$= 220 MeV on the $L=32$ lattice (open squares).
The filled symbols are those of $\chi_A^{\rm con.}$ and $\chi_A^{\rm dis.}$,
which dominate the signals.
The other contributions $\Sigma_{\rm sub.}(m)/m$ (circles) and the $SU(2)$ susceptibilities $\Delta_{SU(2)}^{(1,2)}(m)$
(circles and triangles) are relatively small.
This result indicates that the connected part of the subtracted chiral susceptibility
is essentially described by
the axial $U(1)$ susceptibility 
and the disconnected susceptibility is governed by the topological susceptibility\footnote{
  It was pointed out in \cite{Buchoff:2013nra, 1826587} that
  $\chi^{\rm dis.}$ is dominated by the $U(1)_A$ breaking in the $m=0$ limit,
  but the concrete form (in terms of the topological susceptibility) at finite $m$
  was not discussed.
}. The axial $U(1)$ breaking contributions $\chi_A^{\rm con.}$ and $\chi_A^{\rm dis.}$
are strongly suppressed at the lightest quark mass.
In the data with different lattice sizes $L=24$ (crosses) and 40 (stars), no significant volume dependence is seen.
The data may indicate a peak at $m=0.005$.

\begin{figure*}[bth]
  \centering
    \includegraphics[width=12cm]{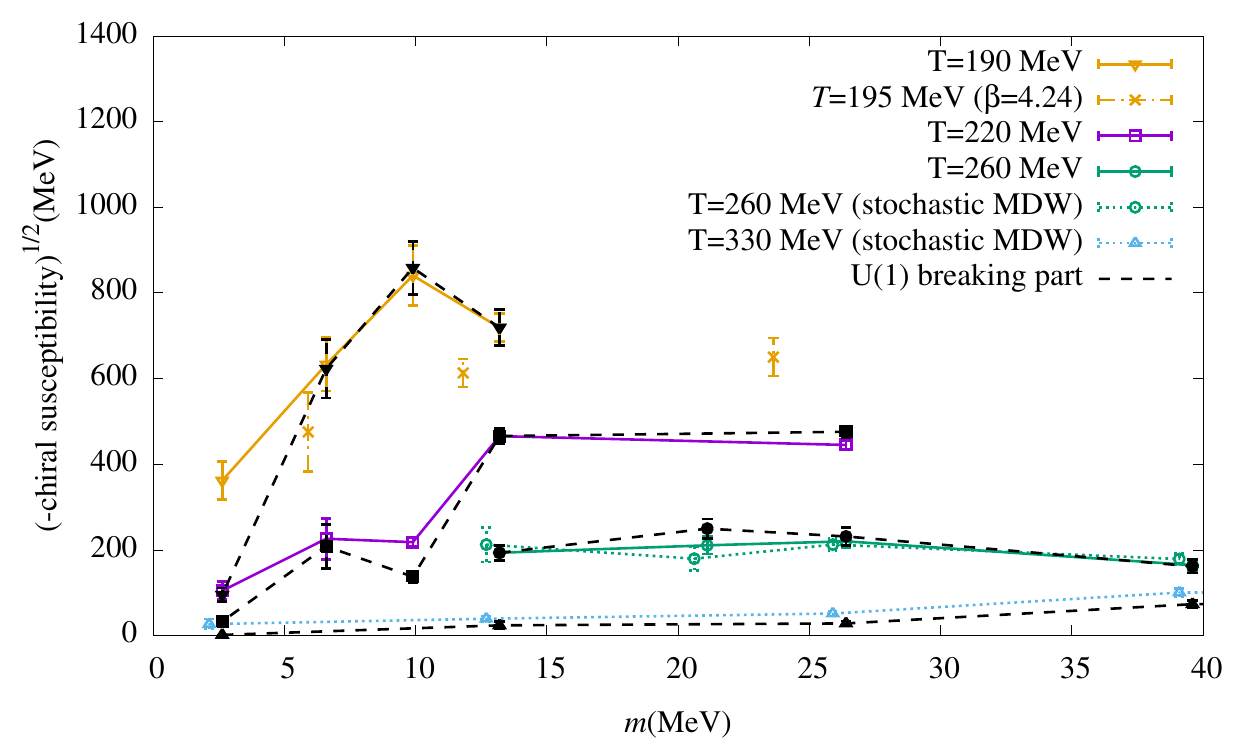}
    \includegraphics[width=12cm]{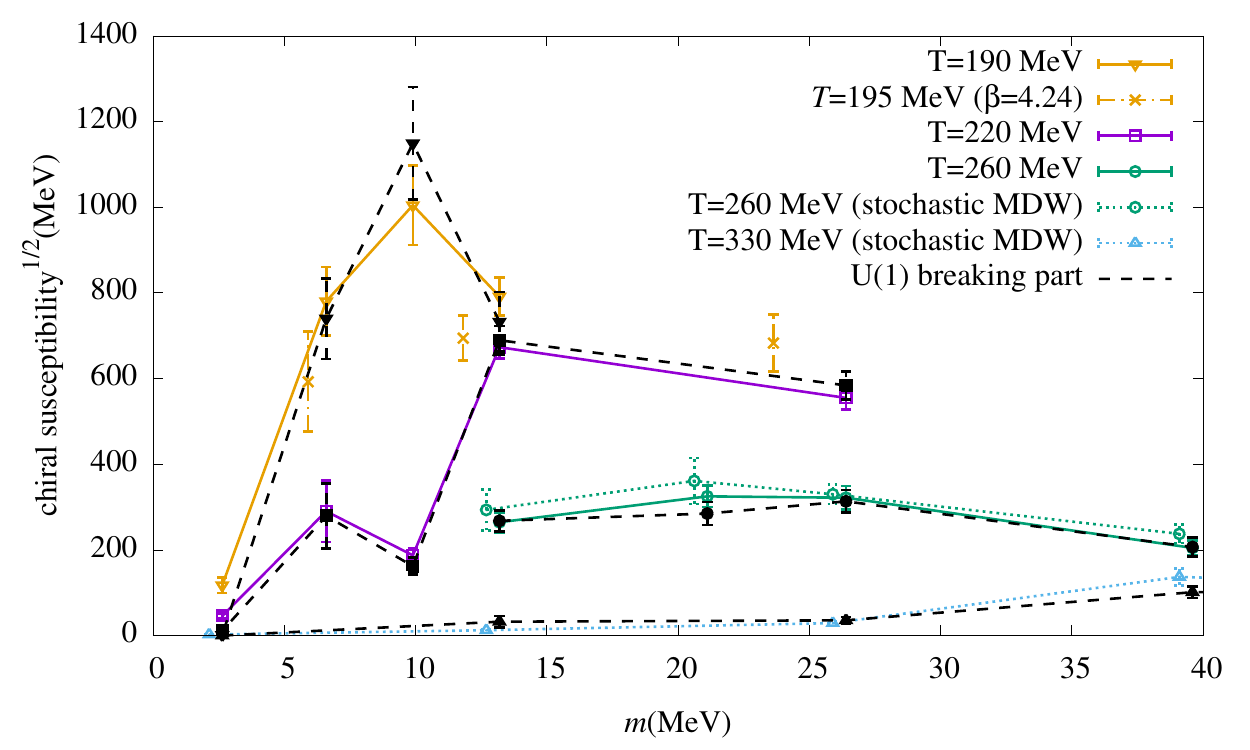}
  \caption{
    Chiral susceptibility at four different temperatures on the $L=32$ lattices (open symbols).
    The connected (top panel) and disconnected (bottom) parts are shown.
    The filled symbols are those from the axial $U(1)$ breaking.
  }
  \label{fig:result}
\end{figure*}

These features are seen at all simulated temperatures
and quark masses ranging from the physical point to $m\sim 100$ MeV.
Figure~\ref{fig:result} summarizes the quark mass dependence of
the connected (top panel) and disconnected (bottom)
chiral susceptibility 
at four different temperatures on the $L=32$ lattices.
The open symbols with solid lines are the data obtained from
the eigenmode decomposition of the reweighted overlap-Dirac operator,
while those with dotted lines are from direct measurement with the
M\"obius domain-wall fermion.
At each temperature, the axial $U(1)$ breaking effect
$\chi_A^{\rm con./dis.}$ (filled symbols with dashed lines)
dominates the signal of the susceptibility.
It is remarkable that this dominance is seen even at higher quark mass region
than the peaks, where both $SU(2)_L\times SU(2)_R$ and $U(1)_A$
are expected to be sizably broken.
In fact, we find that a simple sums of $\chi^{\rm con.}_A$ and $\chi^{\rm dis.}_A$
over 26 simulated data points are $-0.48(2)$ and $0.74(6)$, respectively,
while those of $\chi^{\rm con.}_{\rm sub.}$ and $\chi^{\rm dis.}$ are
$-0.50(2)$ and $0.73(4)$. They differ by only 3\% and 1\% and within standard deviation.
Also, we note that the axial $U(1)$ breaking contributions are strongly
suppressed near the chiral limit.

We also plot the result obtained in our previous work
with $\beta=4.24$ on a coarser lattice ($a$=0.084 fm)
at $T=195$ MeV (cross symbols). The result is consistent with
our new data at a similar temperature $T=190$ MeV,
which indicates that the cut-off effect is not significant.

In Fig.~\ref{fig:result} the position of the  peak moves
towards heavier quark masses
as temperature increases. It indicates that
our simulated temperatures cover
the pseudo-critical temperature,
which becomes higher for larger quark masses\footnote{
  A strong increase of the pseudo-critical temperature
  was reported in 2+1-flavor lattice QCD simulations \cite{Ding:2019prx}.
}.

\begin{figure*}[bth]
  \centering
    \includegraphics[width=12cm]{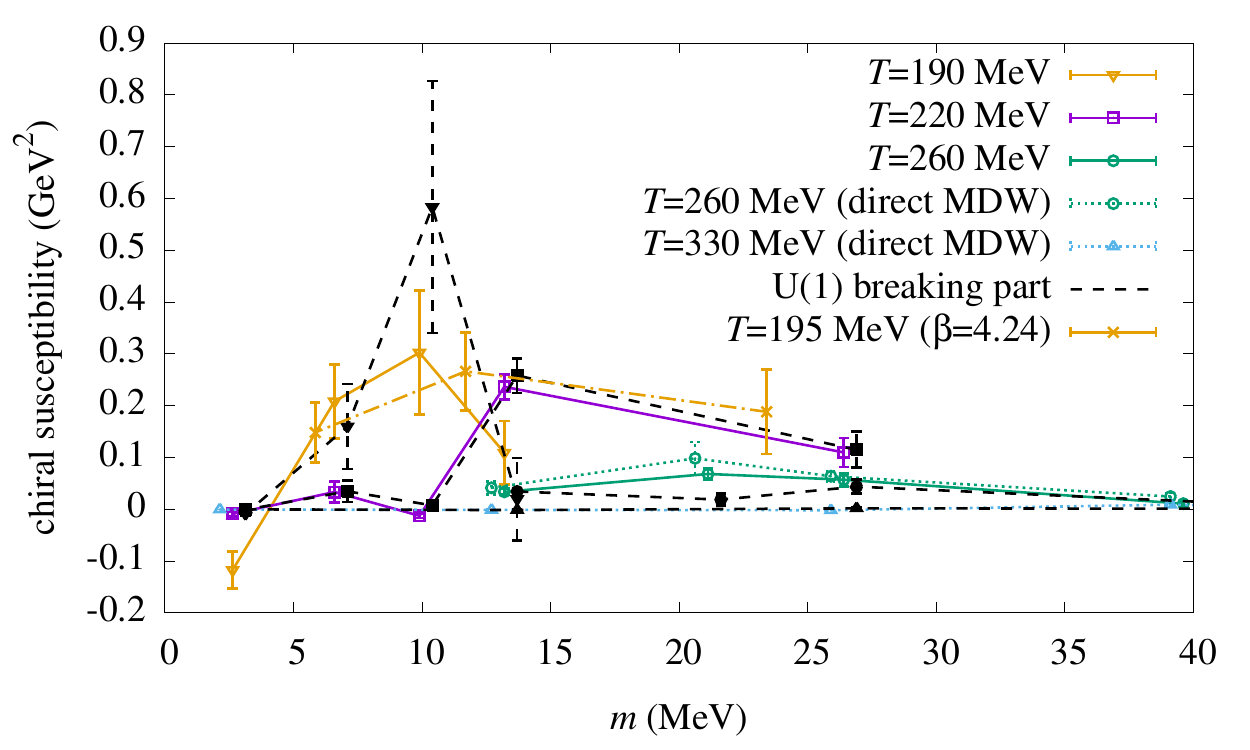}
    \includegraphics[width=12cm]{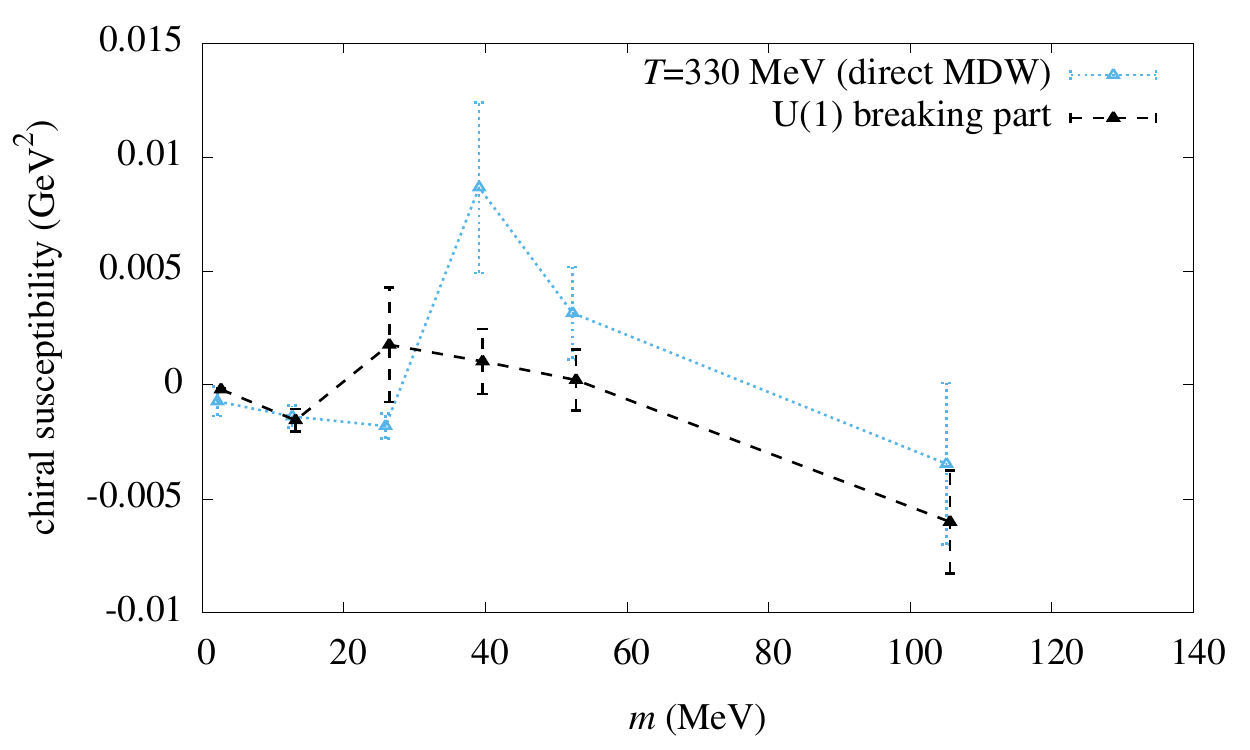}
  \caption{
    Chiral susceptibility at four different temperatures on the $L=32$ lattices (open symbols).
    The filled symbols are those from the axial $U(1)$ breaking.
    The bottom panel is the same plot as the top but
    the result at $T=330$ MeV is shown in a fine scale.
  }
  \label{fig:result2}
\end{figure*}
\begin{figure*}[bth]
  \centering
    \includegraphics[width=12cm]{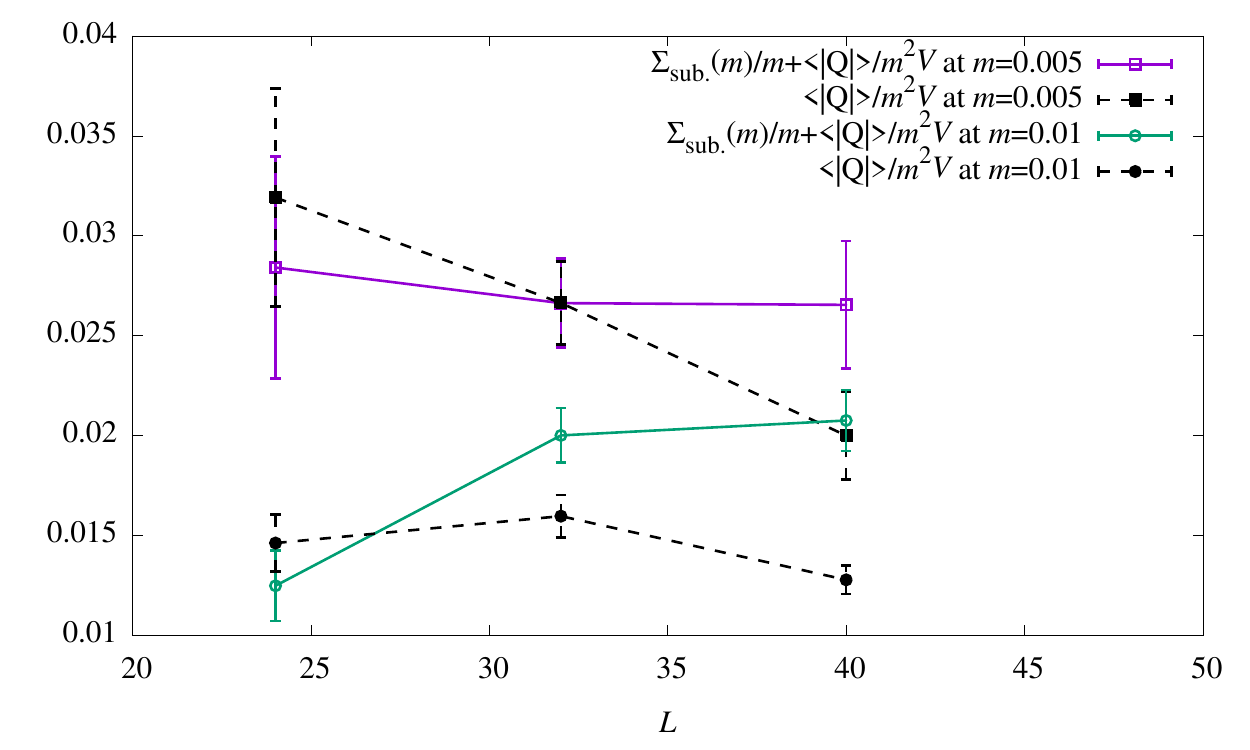}
  \caption{
    The lattice size $L$ dependence of $\frac{\Sigma_{\rm sub.}(m)}{m}+\frac{\langle |Q(A)|\rangle}{m^2V}$
    (open symbols connected by solid lines)
    and $\frac{\langle |Q(A)|\rangle}{m^2V}$ (filled symbols connected by dashed lines). The data at $m=0.01$ and $0.005$ are presented.
  }
  \label{fig:Vdep}
\end{figure*}

In the total contribution  $\chi^{\rm con.}_{\rm sub.}(m)+ \chi^{\rm dis.}(m)$, however,
the situation is not so simple.
As shown in Fig.~\ref{fig:result2}, the $U(1)_A$ breaking dominance by
\begin{equation}
  \chi^{\rm con.}_A(m)+ \chi^{\rm dis.}_A(m) = -\Delta_{U(1)}^{\rm con.}(m)+ \frac{N_f}{m^2}\chi_t(m)+\frac{\langle |Q(A)|\rangle}{m^2V} ,
\end{equation}
is still visible.
But the smallness of $\Delta_{SU(2)}^{(1),(2)}(m)$ implies $\Delta_{U(1)}^{\rm con.}(m) \sim \frac{N_f}{m^2}\chi_t(m)$
so that the quantity is dominated by the last term $\frac{\langle |Q(A)|\rangle}{m^2V}$,
which is expected to vanish in the thermodynamical limit.
Therefore, in order to quantify the axial $U(1)$ breaking effect in the total contribution,
we need a careful analysis of the delicate cancellation between $\Delta_{U(1)}^{\rm con.}(m)$ and $\frac{N_f}{m^2}\chi_t(m)$,
as well as their large volume limits.
Although such a fine analysis is beyond the scope of this work,
let us try to raise two possible scenarios. The first one is that
the signal of the total susceptibility
gets smaller as the volume increases and it is eventually given by
the tiny quark mass dependence of the $SU(2)_L\times SU(2)_R$ breaking.
The second is that even when $\frac{\langle |Q(A)|\rangle}{m^2V}$ disappears
in the large volume limit, the near-chiral-zero modes in $\Sigma_{\rm sub.}(m)$ compensate its absence
and keep the total susceptibility insensitive to the volume. 
In Fig.~\ref{fig:Vdep}, we plot $\frac{\Sigma_{\rm sub.}(m)}{m}+\frac{\langle |Q(A)|\rangle}{m^2V}$
(open and solid symbols) and $\frac{\langle |Q(A)|\rangle}{m^2V}$ (filled and dashed)
as functions of the lattice size $L$.
The consistency of the former at $L=32$ and 40 in spite of the decrease of the latter
may be a support of the second scenario.

We conclude that the connected and disconnected chiral susceptibilities
are dominated by the axial
$U(1)$ breaking effects at temperatures $T\gtrsim 190$ MeV, which
covers the pseudo-critical temperature when the quark mass is finite.
The connected part 
is described by the axial $U(1)$ susceptibility other than the $m$-independent quadratically divergent part,
and the disconnected part is governed by the topological susceptibility.
The chiral limit of the $U(1)_A$ contributions is strongly suppressed.
The picture of QCD phase diagram \cite{Pisarski:1983ms}, based on 
the spontaneous $SU(2)_L\times SU(2)_R$ breaking alone, may need to be reconsidered,
or a delicate cancellation between the connected and disconnected parts of the
$U(1)_A$ breaking is, at least, required.
It is turned out that the axial $U(1)$ breaking does play a crucial role.




\section*{Acknowledgment}
We thank 
H.-T. Ding, C. Gattringer, L. Glozman,
for useful discussions. We thank P. Boyle for correspondence for starting simulation with Grid 
and I. Kanamori for helping us on the simulations on K computer with Bridge++.
We also thank the members of JLQCD collaboration for their encouragement and support.
We thank the Yukawa Institute for Theoretical Physics at Kyoto University.
Discussions during the YITP workshop YITP-W-20-08 on "Progress in Particle Physics 2020" were useful to complete this work.
Numerical simulations were performed using the QCD software packages Iroiro++ \cite{Cossu:2013ola},
   Grid \cite{Boyle:2015tjk}, and Bridge++ \cite{Ueda:2014rya}
on IBM System Blue Gene Solution at KEK under a
support of its Large Scale Simulation Program (No. 16/17-14) and Oakforest-PACS at JCAHPC
under a support of the HPCI System Research Projects (Project IDs: hp170061, hp180061,
hp190090, and hp200086), Multidisciplinary Cooperative Research Program in CCS, University of Tsukuba
(Project IDs: xg17i032 and xg18i023) and K computer provided by the RIKEN Center for Computational Science.
We used Japan Lattice Data Grid (JLDG) \cite{Amagasa:2015zwb}
for storing a part of the numerical data generated for this work.
This work is supported in part by the Japanese Grant-in-Aid for Scientific Research
(No. JP26247043, JP16H03978, JP18H01216, JP18H03710, JP18H04484, JP18H05236), and by MEXT as
“Priority Issue on Post-K computer" (Elucidation of the Fundamental Laws and Evolution of the
Universe) and by Joint Institute for Computational Fundamental Science (JICFuS).


\appendix

\section{Comparison with Ding {\it et al.} \cite{1826587}}

Recently, Ding {\it et al.} \cite{1826587} investigated the disconnected part of
chiral susceptibility in $N_f=2+1$ QCD using eigenvalues
of the Dirac operator of highly improved staggered quark (HISQ) action.
At $T=207$ MeV they obtained a non-zero
continuum limit, which suggests that the axial $U(1)$ symmetry
is still broken by anomaly at 1.6 $T_c$. 
As their conclusion qualitatively differs from ours, which becomes consistent with zero
at the lightest simulated quark mass, here we would like to
compare the two. 

\begin{figure}[bth]
  \centering
    \includegraphics[width=12cm]{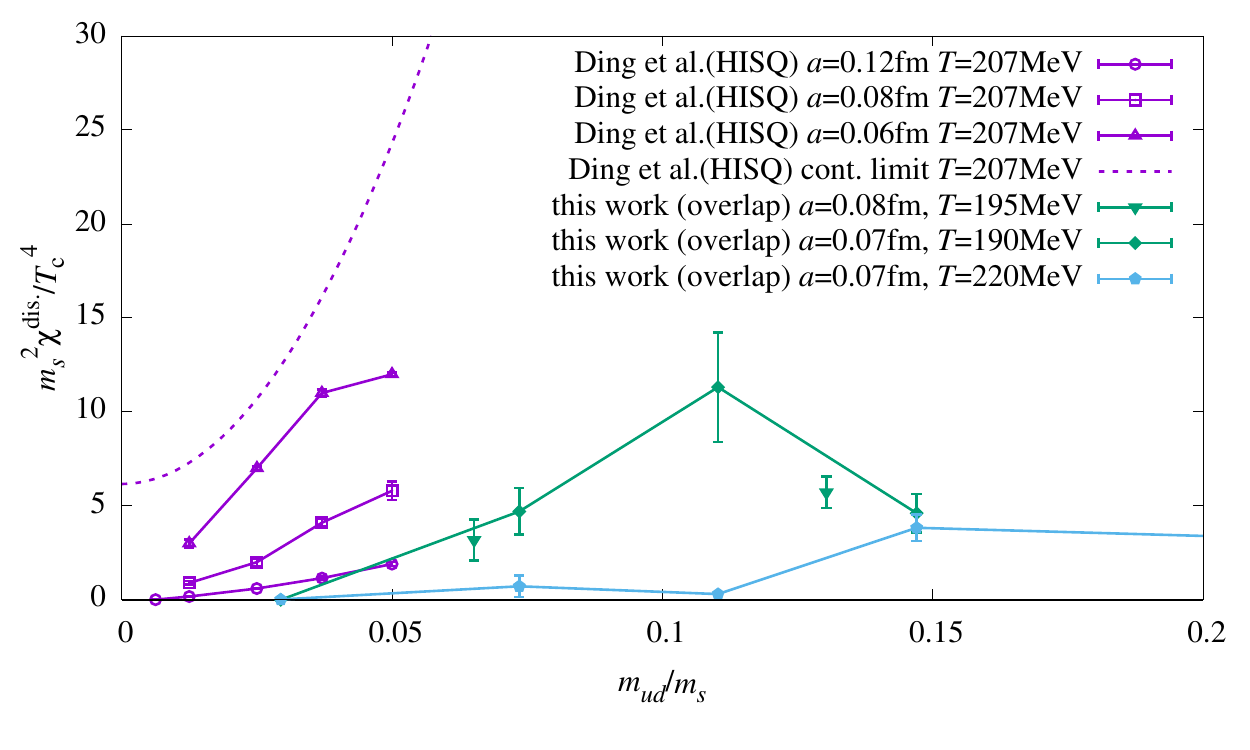}
  \caption{
    Comparison of $\chi^{\rm dis.}(m)$ between Ding {\it et al.} \cite{1826587}
    (open symbols) and this work (filled).
    A qualitative feature of the sharp drop towards the chiral limit is similar.
    But the large scaling violation in \cite{1826587} leads to
    a continuum limit much larger than the raw values as shown in the dashed curve.
  }
  \label{fig:compare}
\end{figure}

In Fig.~\ref{fig:compare} we present
the data of \cite{1826587} with open symbols
and that of this work with the filled symbols.
Since the strange quark is quenched in our simulations,
we simply use the physical value of the strange quark mass for $m_s$.
Note that the critical temperature is estimated to be $\sim$ 130 MeV for $N_f=2+1$
QCD while it is $\sim170$ MeV for $N_f=2$.
Interestingly, a qualitative feature of sharp drops
towards the chiral limit is similar.
However, a significant cutoff $1/a$ dependence is seen in \cite{1826587}
the data at $a=0.06$ fm are twice larger than those at $a=0.08$ fm,
while our data 
at $a=0.08$ fm ($T=195$ MeV) and those
at $a=0.07$ fm ($T=190$ MeV)
do not show such a sizable discretization effect.

In \cite{1826587} they obtained a continuum limit
with a global fit with 6 parameters, which is shown  by the dashed curve in
Fig.~\ref{fig:compare}. It is clearly higher than the raw data at finite lattice spacings.
Specifically, the one at the second lightest quark mass at $a=0.12$ fm
is extrapolated to a continuum limit that is 40 times larger,
which suggests that their lattice data are not on a proper scaling trajectory that allows
continuum extrapolation assuming an expansion in $a^2$.
Since the $U(1)_A$ anomaly does not correctly couple to the taste singlet component of
the staggered fermion that \cite{1826587} employed,
the quantities which are highly affected by the chiral anomaly and index theorem
may receive large discretization effects in their simulation.


\end{document}